\documentclass[prb]{revtex4}
\usepackage{graphicx,epsfig}

\begin{document}

\title{Two-dimensional metal-insulator transition and in-plane magnetoresistance in a high mobility strained $Si$ quantum well}

\author{K. Lai$^1$, W. Pan$^2$, D.C. Tsui$^1$, S.A. Lyon$^1$, M. M$\ddot{u}$hlberger$^3$, and F. Sch$\ddot{a}$ffler$^3$}

\affiliation{$^1$ Department of Electrical Engineering, Princeton University, Princeton, NJ 08544}

\affiliation{$^2$ Sandia National Laboratories, P.O. Box 5800, MS 0601, Albuquerque, NM 87185}

\affiliation{$^3$ Universit$\ddot{a}$t Linz, Institut f$\ddot{u}$r Halbleiterphysik, Linz, Austria}

\vskip5pc

\begin{abstract}

The apparent metal-insulator transition is observed in a high quality two-dimensional electron system (2DES) in the strained
$Si$ quantum well of a $Si/Si_{1-x}Ge_x$ heterostructure with mobility $\mu=1.9\times10^5$ cm$^2$/Vs at density
$n=1.45\times10^{11}$ cm$^{-2}$. The critical density, at which the thermal coefficient of low $T$ resistivity changes sign,
is $\sim 0.32\times10^{11}$ cm$^{-2}$, so far the lowest observed in the $Si$ 2D systems. In-plane magnetoresistance study
was carried out in the higher density range where the 2DES shows the metallic-like behavior. It is observed that the in-plane
magnetoresistance first increases as $\sim B_{ip}^2$ and then saturates to a finite value $\rho(B_c)$ for $B_{ip} > B_c$. The
full spin-polarization field $B_c$ decreases monotonically with $n$ but appears to saturate to a finite value as $n$
approaches zero. Furthermore, $\rho(B_c)/\rho(0) \sim 1.8$ for all the densities ranging from $0.35\times10^{11}$ to
$1.45\times10^{11}$ cm$^{-2}$ and, when plotted versus $B_{ip}/B_c$, collapses onto a single curve.

\end{abstract}


\date{\today}
\maketitle

The two-dimensional (2D) metal-to-insulator transition (MIT) has been of great current interests.
\cite{kravchenko04,pudalov03} According to the well-established scaling theory, \cite{abrahams79} any amount of disorder in a
non-interacting 2D electron system (2DES) will localize the carriers at zero temperature ($T$) and zero magnetic ($B$) field,
and the ground state of the 2DES is an insulator, whose conductivity logarithmically goes to zero as $T \to 0$. Recent
experimental studies in high quality dilute 2D systems where the electron-electron ($e-e$) interaction is large, however,
have shown the existence of a metallic-like state and an apparent metal-insulator transition. There, the magnitude of the
2DES resistivity ($\rho$) undergoes a change from decreasing with decreasing $T$, ($d\rho/dT > 0$), a metallic behavior, to
increasing with decreasing $T$, ($d\rho/dT < 0$), an insulating behavior, at a critical density $n_c$.

One of the fundamental questions in this apparent MIT problem is the nature of the metallic state \cite{dassarma04} and its
response to a pure in-plane magnetic field ($B_{ip}$).
\cite{simonian97,yoon99,okamoto99,vitkalov00,tutuc01,shashkin01,pudalov02,zhu03,lewalle04} It is observed that in
conventional clean Si-MOSFET's the in-plane magnetoresistance (MR) of the 2DES, $\rho(B_{ip})$, first increases as $B_{ip}^2$
at low $B_{ip}$. After a critical B field $B_c$, which has been identified as the full spin polarization B field for the
2DES, \cite{okamoto99,vitkalov00,tutuc01} the in-plane MR saturates to a constant value $\rho(B_c)$. \cite{note} The
enhancement of $\rho(B_{ip})$ under high in-plane B field in this 2DES is attributed to the reduction of screening of charged
impurities in a Fermi liquid, caused by the loss of spin degeneracy, \cite{dolgopolov00,herbut01,gold03} and a ratio of
$\rho(B_c)/\rho(0) =4$ or $\sim 1.2$ is expected when the background impurity scattering \cite{dolgopolov00} or the remote
ionized impurity scattering \cite{gold03} dominates. Thus, it is surprising that close to the critical density $n_c$, the
enhancement can be as large as several orders of magnitude.
\cite{kravchenko04,pudalov03,simonian97,okamoto99,vitkalov00,shashkin01,pudalov02} Furthermore, it has been shown that under
$B_{ip}$ the metallic state is suppressed, and completely destroyed for $n < 1.5n_c$. A recent study has demonstrated that
the disappearance of the positive $d\rho/dT$ in large $B_{ip}$ is due to a competition between weak localization and other
mechanisms, such as screening. \cite{lewalle04}

So far, most of the large number of experiments on 2D MIT in the $Si$ 2DES are carried out using the so-called clean
Si-MOSFET structures with peak electron mobility $\mu$ of $\sim 4\times10^4$ cm$^2$/Vs. In recent years, the 2DES in the high
quality strained $Si$ quantum well (QW) in $Si/SiGe$ heterostructures has emerged as a promising $Si$ system to study strong
electron-electron interaction physics, $e.g.$, the fractional quantum Hall effect. \cite{lai04} Due to modulation doping and
smooth interface, electron mobility at least 2-3 times better than that in the cleanest Si-MOSFET's, can be routinely
achieved.

In this communication, we report experimental results on the apparent metal-insulator transition in a high quality 2DES in
the strained $Si$ quantum well of a $Si/Si_{1-x}Ge_x$ heterostructure. The critical density is found to be $\sim
0.32\times10^{11}$ cm$^{-2}$, much smaller than that observed in clean Si-MOSFET's, where $n_c \sim 0.8\times10^{11}$
cm$^{-2}$. The in-plane magnetoresistivity ($\rho(B_{ip})$) measurements were carried out in the density regime where the
2DES shows the metallic-like behavior at $B=0$. It is observed that $\rho(B_{ip})$ first increases as $\sim B_{ip}^2$ and
then saturates to a finite value $\rho(B_c)$ for $B > B_c$. The full spin-polarization field $B_c$ decreases monotonically
with $n$ but appears to saturate to a finite value as $n$ approaches zero. Furthermore, $\rho(B_c)/\rho(0) \sim 1.8$ for all
the densities ranging from $0.35\times10^{11}$ to $1.45\times10^{11}$ cm$^{-2}$ and, when plotted versus $B_{ip}/B_c$,
collapses onto a single curve.

The starting specimen is an MBE-grown $Si/S_{1-x}Ge_x$ heterostructure, with a 15 nm wide strained $Si$ quantum well. Details
of the growth structure is given in Ref. [19]. A field-effect transistor type device was then fabricated. \cite{lai04b} At $T
\sim 300$~mK and zero gate voltage, the 2DES has a density $n=1.45\times10^{11}$ cm$^{-2}$ and mobility $\mu=190,000$
cm$^2$/Vs. Standard low-frequency ($\sim$~7Hz) lock-in techniques were used to measure the 2D transport coefficients.

We show in Figure 1 the magnetoresistivity $\rho_{xx}$ and Hall resistance $\rho_{xy}$ at $n=0.42\times10^{11}$ cm$^{-2}$.
The appearance of strong integer quantum Hall effect (IQHE) states at Landau level fillings $\nu=1,2$, as well as at $\nu =
4$ demonstrate high quality of the 2DES.

In Figure 2, we show the temperature dependence of $\rho$, the zero $B$ resistivity, at selected densities. At $n \geq
0.63\times10^{11}$ cm$^{-2}$, $\rho$ monotonically decreases with decreasing temperature. Below $0.3\times10^{11}$ cm$^{-2}$,
$\rho$ increases rapidly with decreasing $T$ and the 2DES is in the insulating regime. The critical density $n_c$, at which
the thermal coefficient of low $T$ resistivity changes sign, is $0.32\times10^{11}$ cm$^{-2}$. We emphasize that it is by far
the lowest $n_c$ obtained in the $Si$ based 2D systems. In the so-called transition regime between $0.35\times10^{11} \leq n
\leq 0.55\times10^{11}$ cm$^{-2}$, $\rho$ first increases with decreasing $T$, reaches a maximum, and then decreases with
continuously decreasing temperature.

One of the fundamental questions in this apparent MIT problem is the in-plane magnetoresistivity $\rho(B_{ip})$.
\cite{simonian97,yoon99,okamoto99,vitkalov00,tutuc01,shashkin01,pudalov02,zhu03,lewalle04} In Figure 3(a), $\rho(B_{ip})$ is
plotted for several densities. Like in Si-MOSFET's, $\rho(B_{ip})$ first increases as $\sim B_{ip}^2$. After a critical $B$
field $B_c$, it saturates to a roughly constant value, $\rho(B_c)$. In Figure 3(b), $\rho(B_{ip})/\rho(0)$ is plotted versus
$B_{ip}/B_c$. It is clear that in this large density range, from $0.35\times10^{11}$ to $1.45\times10^{11}$ cm$^{-2}$,
$\rho(B_{ip})/\rho(0)$ collapse onto a single curve, and $\rho(B_c)/\rho(0) \sim 1.8$ for all the densities.

$B_c$ in Figure 3(a) represents the $B$ field beyond which electrons become fully spin-polarized.
\cite{okamoto99,vitkalov00,tutuc01} In Figure 3(c), we plot $B_c$ as a function of $n$. We also include the data obtained by
Okamoto {\it et al}, \cite{okamoto04} measured at densities $n
> 1 \times 10^{11}$ cm$^{-2}$. Results from two experiments are in good agreement with each other. At high $n$'s, $B_c$
decreases roughly linearly with $n$. In fact, for $n > 0.8 \times 10^{11}$ cm$^{-2}$, $B_c = -1.38 + 5.55 \times n$. The
decreasing rate slows down at lower electron densities and deviates from that of the linear dependence. $B_c$ appears to
approach a finite value as $n \to 0$.

In Figure 4, we show the temperature dependence of the in-plane MR at three different densities. At the high density of
$n=0.515\times10^{11}$ cm$^{-2}$, the 2DES remains metallic even at $B_{ip}=7$~T, much higher than the critical $B$ field of
$B_c \sim 2$~T. At the intermediate density of $n=0.38\times10^{11}$ cm$^{-2}$, the 2DES shows metallic behavior at small
$B_{ip}$, becomes insulating at $B_{ip} \sim 1.5$~T, and then re-enters into the metallic phase at higher $B_{ip}$. When $n$
is further reduced to $n=0.35\times10^{11}$ cm$^{-2}$, the in-plane field simply destroys the zero $B$ metallic phase, and
the 2DES becomes insulating at $B_{ip} > ~\sim 1$~T.

The observed critical density $n_c=0.32\times10^{11}$ cm$^{-2}$ is about one order of magnitude smaller than the $n_c$
observed in lower quality $SiGe$ systems (for example, $n_c = 2.35\times10^{11}$ cm$^{-2}$ in Ref.~[20] with the highest 2DES
mobility of $\mu = 7.5\times10^4$ cm$^2$/Vs, and $n_c=4.05\times10^{11}$ cm$^{-2}$ in Ref.~[22] with the highest mobility
$\mu = 6.0 \times10^4$ cm$^2$/Vs) and $\sim$ 2-3 times smaller than that in clean Si-MOSFET's, where $n_c \sim
0.8\times10^{11}$ cm$^{-2}$. \cite{kravchenko04} Thus, our result demonstrates that $n_c$ in the $Si$ systems also decreases
as the sample quality increases, consistent with previous observations in the $GaAs$ systems. \cite{yoon99b} Furthermore, at
the critical density of $n_c=0.32\times10^{11}$ cm$^{-2}$, the dimensionless $e-e$ interaction parameter $r_s =
(\pi/n)^{1/2}(e/h)^2(m^*/\epsilon\epsilon_0)$ is 10, where $m^*=0.2m_e$ is the electron band mass and $\epsilon=11.7$ is the
dielectric constant for strained $Si$, and other parameters have their usual meanings. This is the largest critical $r_s$ so
far obtained in the $Si$ based 2DES's. We note here that this number, however, is only slightly higher than that ($r_s \sim
9.3$ at $n_c=0.8\times10^{11}$ cm$^{-2}$) reported in clean Si-MOSFET's. But, as pointed by Okamoto {\it et al},
\cite{okamoto04} in previous calculations in clean Si-MOSFET's, the critical $r_s$ value was over estimated due to the use of
an average relative dielectric constant of silicon and $SiO_2$, $\epsilon_{av}=7.7$, in the calculation of $r_s =
(\pi/n)^{1/2}(e/h)^2(m^*/\epsilon_{av}\epsilon_0)$. Considering that the average distance of the 2D electrons from the
$Si/SiO_2$ interface is comparable to the average distance between two electrons, the realistic $\epsilon$ should be larger
than 7.7, \cite{okamoto04} thus reducing the critical $r_s$ to $\sim 8$ in clean Si-MOSFET's.

The observed enhancement of the $\rho(B_{ip}$) under high in-plane $B$ field can be explained by the reduction of screening
of charged impurities in a Fermi liquid, caused by the loss of spin degeneracy. \cite{dolgopolov00,herbut01,gold03} It has
been shown that, when the background impurity scattering dominates, a ratio of $\rho(B_c)/\rho(0) =4$ is expected.
\cite{dolgopolov00} On the other hand, when the remote ionized impurity scattering prevails, this ratio is reduced to  $\sim
1.2$. \cite{gold03} The ratio of $\rho(B_c)/\rho(0) \sim 1.8$ in our measurements sits between these two limits and is closer
to 1.2, indicating that the dominating scattering mechanism at low temperatures is from remote ionized impurities. This is
consistent with our sample growth structure, where the doping layer is 150 $\AA$ away from the 2D electron channel. The
slightly higher ratio than 1.2 is probably related to the strain field, which can act as background scattering centers. What
is surprising in our results is that in a wide density range the enhancement of $\rho(B_c)/\rho(0)$ is the same, even at the
density of $n=0.35\times10^{11}$ cm$^{-2}$, which is only 10\% higher than the critical density of $n_c=0.32\times10^{11}$
cm$^{-2}$. On the contrary, in high quality Si-MOSFET's, while $\rho(B_c)/\rho(0) \sim 2.2$ at very high densities,
\cite{broto03} the enhancement is much higher when $n$ is close to $n_c$, {\it e.g.}, $\rho(B_c)/\rho(0) \sim 10$ at
$0.89\times10^{11}$ cm$^{-2}$, \cite{vitkalov00} which also is about 10\% higher than the $n_c$ of $\sim 0.8\times10^{11}$
cm$^{-2}$. We speculate that this quantitative difference in $\rho(B_c)/\rho(0)$ as $n \to n_c$ in the two systems is
probably related to a smoother interface between $Si$ and $SiGe$ and thus, less surface roughness scattering, in our high
quality strained $Si$ quantum well. Finally, we note that a similar enhancement of $\rho(B_c)/\rho(0) \sim 1.7$ was also
observed in other high quality strained $Si$ QW samples. \cite{okamoto04,dolgopolov03}

In summary, in a high electron mobility 2DES realized in the strained $Si$ quantum well, we observe that, with increasing
sample quality, the critical density in the 2D metal-insulator transition decreases to a smaller value. Moreover, the
measured full spin-polarization magnetic field, $B_c$, decreases monotonically with $n$ but appears to saturate to a finite
value as $n$ approaches zero. The saturation value of the in-plane magnetoresistivity, $\rho(B_c)$, over the zero field
resistivity is constant, $\sim 1.8$, for all the densities ranging from $0.35\times10^{11}$ to $1.45\times10^{11}$ cm$^{-2}$
and, when plotted versus $B_{ip}/B_c$, $\rho(B_{ip})/\rho(0)$ collapses onto a single curve.

We would like to thank G. Cs$\acute{a}$thy for technical help and S.V. Kravchenko for discussion. The work at Princeton was
supported by AFOSR under grant No. F49620-02-1-0179 and the NSF DMR0352533. Sandia National Laboratories is a multiprogram
laboratory operated by Sandia Corporation, a Lockheed-Martin company, for the U.S. Department of Energy under Contract No.
DE-AC04-94AL85000.
\\
\\
Note added: E.H. Hwang and S. Das Sarma have calculated the temperature, density, and parallel magnetic field dependence of
low temperature 2D resistivity based on our sample structure and that of Okamoto {\it et al}. They show in the following
theoretical paper that the results observed in this paper can be qualitatively and for some parts of the data
semi-quantitatively reproduced using a model where the 2D carriers are scattered by screened random Coulombic impurity
centers.

\begin{figure} [h]
\centerline{\epsfig{file=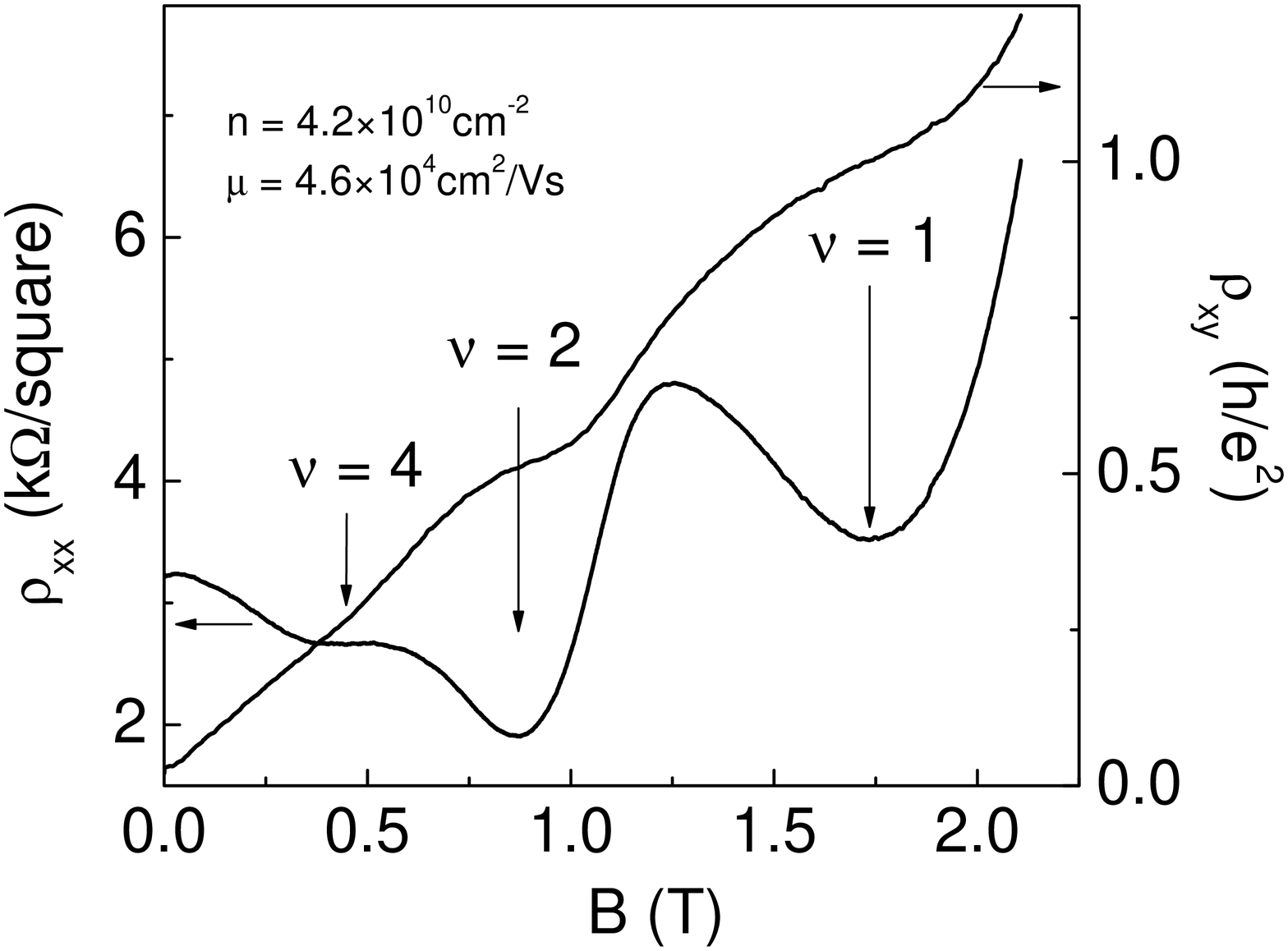,width=6.5cm}}
\caption{The magnetoresistivity and the Hall resistance at the density of
$n=0.42\times10^{11}$ cm$^{-2}$. The sample temperature is 0.3~K. The vertical arrows mark the positions of the IQHE states.}
\end{figure}

\begin{figure} [h]
\centerline{\epsfig{file=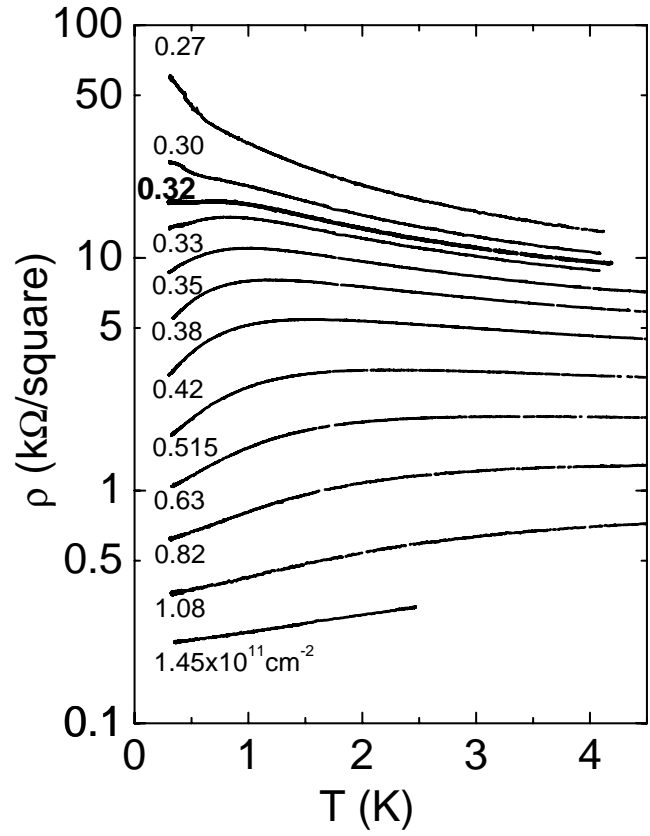,width=10.5cm}}
\caption{ 2D resistivity $\rho$ as a function of temperature at various
densities.}
\end{figure}

\begin{figure} [h]
\centerline{\epsfig{file=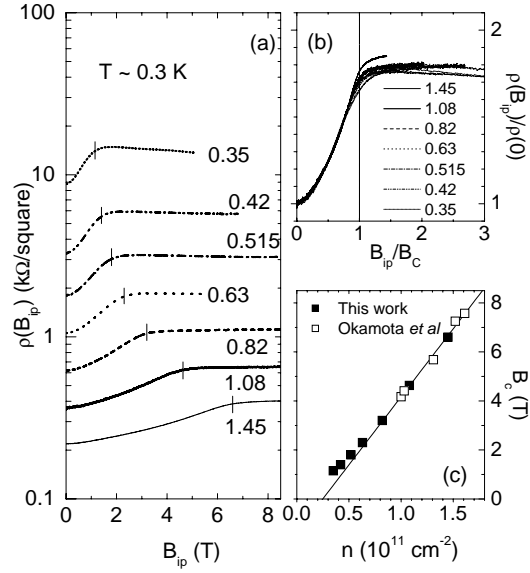,width=7.5cm}} \caption{(a) In-plane magnetoresistivity at a few selected selective 2DES
densities. n is in units of $10^{11}$ cm$^{-2}$. The positions of full polarization $B$ field are marked by the short lines.
(b) The normalized in-plane magnetoresistivity $\rho(B_{ip})/\rho(0)$ vs $B_{ip}/B_c$. (c) $B_c$ as a function of electron
density. Results from Ref. [21] are included. The straight line is a linear fit for densities $n > 0.8 \times 10^{11}$
cm$^{-2}$.}
\end{figure}

\begin{figure} [h]
\centerline{\epsfig{file=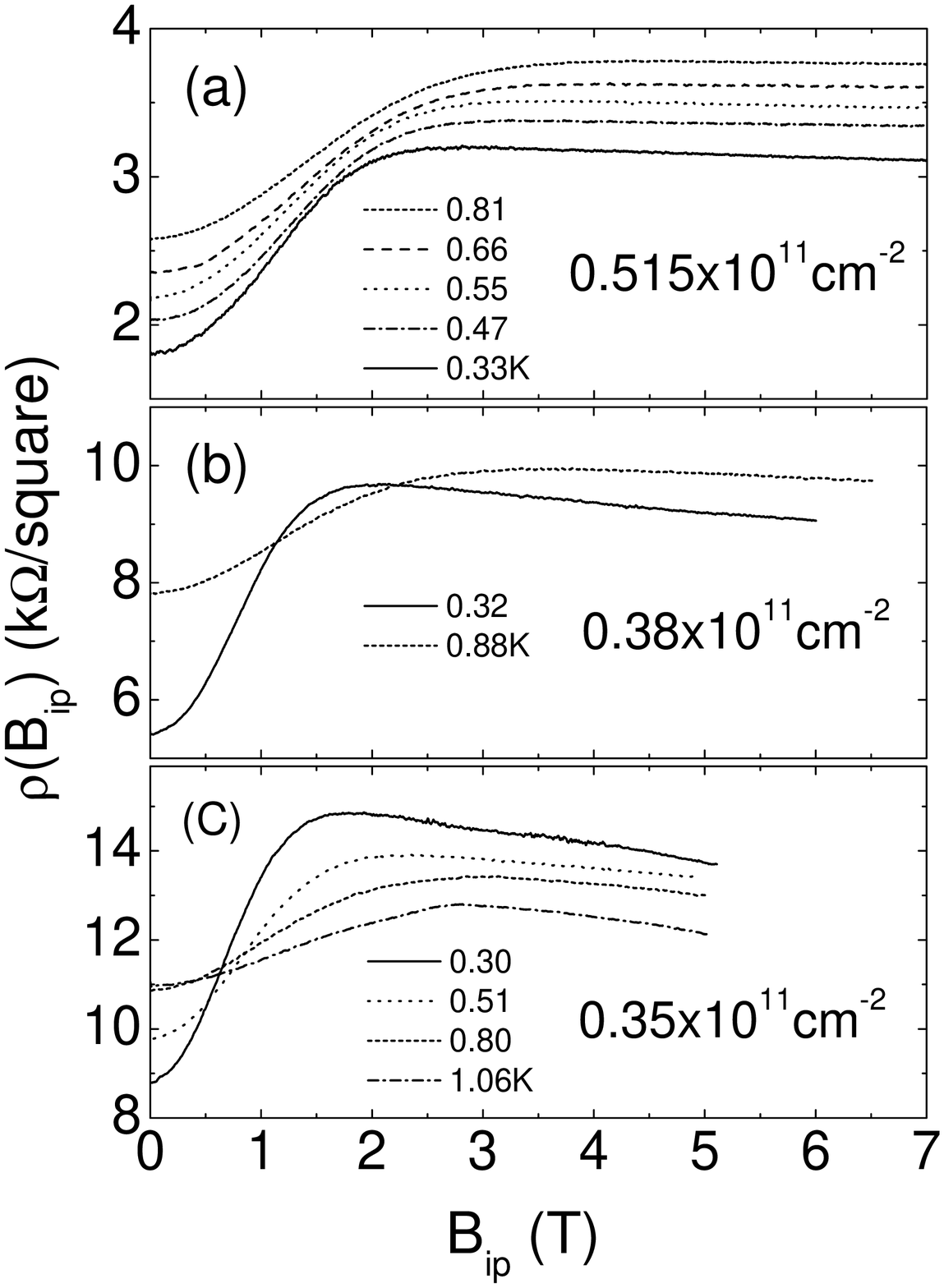,width=7.5cm}} \caption{Temperature dependence of in-plane magnetoresistivity $\rho(B_{ip})$
at three 2DES densities; (a) $n=0.515\times10^{11}$ cm$^{-2}$, (b) $n=0.38\times10^{11}$ cm$^{-2}$, and (c)
$n=0.35\times10^{11}$ cm$^{-2}$.}
\end{figure}


\begin{thebibliography} {90}
\bibitem{note}
In $GaAs$ based systems where the finite width of 2DES is not small, $\rho(B_{ip})$ first increases as exp($B_{ip}^2$) at low
$B_{ip}$ and then continues to increase as exp($B_{ip}$) for  $B_{ip} > B_c$.
\end{thebibliography}
\end{document}